\documentclass[manuscript]{aastex}
\usepackage{rotating}

\newcommand{\Teff}{\mbox{$T_{\rm eff}$}}
\newcommand{\logg}{\mbox{$\log g$}}

\newcommand{\vsini}{\mbox{$v\sin i$}}
\newcommand{\kms}{\mbox{km s$^{-1}$}}

\slugcomment{DRAFT}

\begin{document}

\title{\bf Magnetic Properties of Young Stars in the TW Hydrae Association}

\author{Hao Yang}
\affil{Department of Physics \& Astronomy, Rice University, 6100
Main St. MS-108, Houston, TX 77005} \email{haoyang@rice.edu}

\author{Christopher M. Johns-Krull\altaffilmark{1}}
\affil{Department of Physics \& Astronomy, Rice University, 6100
Main St. MS-108, Houston, TX 77005} \email{cmj@rice.edu}

\author{Jeff A. Valenti\altaffilmark{1}}
\affil{Space Telescope Science Institute, 3700 San Martin Dr.,
Baltimore, MD 21210} \email{valenti@stsci.edu}

\altaffiltext{1}{Visiting Astronomer, Infrared Telescope Facility,
operated for NASA by the University of Hawaii.}

\begin{abstract}

We present an analysis of infrared (IR) echelle spectra of five stars
in the TW Hydrae Association (TWA). We model the Zeeman broadening
in four magnetic-sensitive \ion{Ti}{1} lines near $2.2\ \mu$m and
measure the value of the photospheric magnetic field averaged over
the surface of each star. To ensure that other broadening
mechanisms are properly taken into account, we also inspect several
magnetically insensitive CO lines near $2.3\ \mu$m and find no
excess broadening above that produced by stellar rotation and
instrumental broadening, providing confidence in the magnetic
interpretation of the width of the \ion{Ti}{1} lines. We then 
utilize our results to test the relationship between stellar magnetic 
flux and X-ray properties and compare the measured fields with 
equipartition field values. Finally, we use our results and recent 
results on a large sample of stars in Taurus to discuss the potential
evolution of magnetic field properties between the age of Taurus 
($\sim$2 Myrs) and the age of TWA ($\sim$10 Myrs). We find that
the average stellar field strength increases with age; however,
the total unsigned magnetic flux decreases as the stars contract
onto the main-sequence.

\end{abstract}


\section{Introduction}

T Tauri stars (TTSs) are young, low-mass stars, which 
are newly formed out of their natal molecular clouds and are often
associated with OB associations and/or nebulosity. The location of 
TTSs above the main sequence in the H-R diagram and their lithium
abundance also reveal their youth. TTSs display substantial
photometric and spectroscopic variability, and their properties 
have been reviewed by \citet{appenzeller1989}, \citet{bertout1989}, 
\citet{basribertout1993} and \citet{menardbertout1999}. Depending 
on whether they are actively accreting, TTSs are divided into two 
categories. Classical T Tauri stars (CTTSs) are still surrounded 
by dusty circumstellar disks with detectable non-stellar emission
in the near-infrared (near-IR). Accretion from the disks onto 
the stars produces many observable features such as strong yet 
variable H$\alpha$ emission, often 
displaying P-Cygni or inverse P-Cygni profile shapes. Naked T 
Tauri stars (NTTSs), on the other hand, show little or no sign 
of accretion and do not appear to have close dusty disks around 
them. It is generally believed that stars evolve through a CTTS 
phase to become a NTTS on the way to the main sequence.

One essential piece of the puzzle to our understanding of TTSs is the
stellar magnetic field. For NTTSs, it is believed that dynamo 
generated magnetic fields are responsible for most of their 
peculiarities, e.g., variable X-ray emission, flares, and cool 
spots. The evolution of CTTSs, especially the interaction 
between the stars and their circumstellar disks, is believed to 
be directly controlled by stellar magnetic fields 
\citep[see review by][]{bouvier2007}. Various magnetospheric 
accretion models \citep{camenzind1990, konigl1991, ccc1993, 
shu1994, paatz1996} generally agree that the disks of CTTSs are 
truncated by stellar magnetic fields near the corotation radius 
and the accreting material flows along the field lines toward 
the polar regions of the central star. The disks of CTTSs are 
the birthplace for planets and the radius at which the magnetic 
field truncates the inner disk may determine where giant planets
are halted on their inward migration in the disk 
\citep{lin1996, eisner2005}. Magnetic fields probably also play 
a key role in driving the stellar winds/jets seen in young stars 
\citep[e.g.,][]{shu1999}. Therefore, a full understanding of the 
early evolution of young stars, which will lead to better 
understanding of both star and planet formation, requires 
detailed theoretical consideration and empirical measurements 
of the stellar magnetic field properties.

Direct detection of stellar magnetic fields is difficult,
with measurements starting to become plentiful only recently
\citep{cmj2007}. One technique, 
which led to the first direct detection of magnetic fields on 
a TTS, Tap 35 \citep{basri1992} and later on other stars 
\citep{guenther1999}, is based on an equivalent width increase 
in moderately saturated magnetically sensitive lines due to 
Zeeman splitting. Unfortunately, this method is particularly sensitive 
to the choice of effective temperature and gravity used when analyzing 
the stellar line measurements. Another method of detecting stellar 
magnetic fields is through spectropolarimetry, which looks for 
net circular polarization in Zeeman-sensitive spectral lines. 
Observed along the direction that is parallel to the magnetic field, 
the Zeeman $\sigma$-components are circularly polarized, with the 
components of opposite helicity split to either side of the 
nominal line wavelength. If the magnetic topology on the stars 
is somewhat organized, there should be a measurable component 
along the line of sight. 
Reliable detections of circular polarization in TTS photospheric
absoprtion lines are rare, with most recent studies finding only
upper limits of $100-200$ G \citep{cmj1999a, daou2006, smirnov2004}.
More recently, significant, but similarly weak, detections have been made
\citep{yang2007, donati2007} and one surprising strong detection 
on BP Tau has recently been published \citep{donati2008}.
On the other hand, net 
polarization is detected in the \ion{He}{1} $5876$ \AA\ emission 
line, first discovered by \citet{cmj1999a} on BP Tau. This 
\ion{He}{1} line is believed to form in the postshock region 
\citep{hartmann1994, edwards1994} where material accretes onto 
the star and the detection on BP Tau yielded a net longitudinal 
field of $2.46 \pm 0.12$ kG in the line formation region. Circular 
polarization in the narrow component of the \ion{He}{1} line 
has now been observed in several CTTSs 
\citep{valenti2004, symington2005, yang2007, donati2007, donati2008}. 
These observations suggest that accretion onto 
CTTSs is indeed controlled by a strong stellar magnetic field.

While spectropolarimetry can provide clear detections of stellar
magnetic fields, it is only sensitive to the net field resulting
from the predominance of one field polarity over the other. As
a result, it can miss the majority of the magnetic flux present
on a star. The most successful approach so far for measuring the 
total magnetic flux has been to measure the Zeeman
broadening of spectral lines in unpolarized light
\citep{robinson1980, saar1988, valenti1995, cmj1996, cmj1999b}. 
For any given Zeeman component, the splitting resulting 
from the magnetic field is
$$\Delta\lambda = {e \over 4\pi m_ec^2} \lambda^2 g B
                = 4.67 \times 10^{-7} \lambda^2 g B \,\,\,\,\,\,
                  \rm m\mbox{\AA}\,\eqno(1)$$
where $g$ is the Land\'e $g$-factor of the transition, $B$ is the
strength of the magnetic field in kG, and $\lambda$ is the
wavelength of the transition in \AA. \citet[][hereafter Paper I]{cmj1999b} 
detected Zeeman broadening of the \ion{Ti}{1} line at $2.2233\ \mu$m 
on the CTTS BP Tau and obtained a field strength of $\bar B = 2.6 \pm
0.3$ kG, averaged over the entire surface of the star. From analysis
of four \ion{Ti}{1} lines near $2.2\ \mu$m, 
\citet[][hereafter Paper II]{cmj2004} also detected a strong 
magnetic field of $2.5 \pm 0.2$ kG on the NTTS Hubble 4. In addition, 
these authors examined several CO lines near $2.3\
\mu$m, which are magnetically insensitive, and found no excess
broadening above that due to rotational and instrumental effects. 
This technique has now been used to measure the fields 
of 16 TTSs \citep{cmj2004,yang2005,cmj2007}.



In the largest study to date of fields on TTSs, \citet{cmj2007} used a
sample of 14 stars to look for correlations between the measured magnetic
field properties and stellar properties that might be important for
dynamo action.  No significant correlations were found and \citet{cmj2007}
speculated that the fields seen in these young stars may be entrained
interstellar fields left over from the star formation process 
\citep[e.g.,][]{tayler1987, moss2003}.
To supply more observational constraints and further investigate 
the magnetic properties of TTSs, we present here Zeeman broadening
measurements of five NTTSs in the TW Hydrae association (TWA): TWA
5A, TWA 7, TWA 8A, TWA 9A and TWA 9B. At $\sim 50$ pc 
\citep{wichmann1998}, TWA is the nearest region of recent star 
formation. \citet{kastner1997} first concluded that a few T Tauri 
stars in the apparent neighborhood of TW Hya are indeed physically 
close to each other, based on similar X-ray emission and lithium 
abundance, and proposed the existence of the TW Hydrae Association. 
Over 20 stars have been proposed as members of TWA (though only 
a few of all proposed TWA members have trigonometric parallax 
confirmation), and a compilation of TWA members can be found in 
\cite{zuckerman2004ar}. The age for TWA is estimated at about $10$ Myrs 
\citep[see discussion in][]{navascues2006}, and it is intriguing 
that these pre-main-sequence stars are not associated with any 
molecular cloud. The origin of the TWA is still under investigation 
\citep[e.g., see][]{makarov2007}.

As described earlier, the magnetic field properties of TW Hya (TWA 1)
have been investisgated by \citet[][hereafter, Paper III]{yang2005}
and \citet{yang2007}. Interstingly, \citet{setiawan2008} report
a massive planet very close to TW Hya, and \citet{huelamo2008} argue that
the observed radial velocity variations of TW Hya are best explained by
a cool spot instead of a hot jupiter. Here, we study the magnetic fields of 5
additional members of TWA, utilizing infrared (IR) spectroscopy to 
measure Zeeman broadening in 4 \ion{Ti}{1} lines in the K band. 
The observations and data reduction are presented in \S\ 2. 
In \S\ 3 we describe our data analysis technique and results, and 
it is followed by a discussion of the results in \S\ 4.

\section{Observations and Data Reduction}


We obtained K-band IR spectra of the TWA stars at the NASA
Infrared Telescope Facility (IRTF) in January 2000. Observations
were made with the CSHELL spectrometer \citep{tokunaga90, greene93}
using a 0\farcs5 slit.  With this slit width CSHELL normally delivers
a spectral resolving power of $R \equiv \lambda /\Delta\lambda >
40,000$.  At the time our observations were taken, the resolution
achieved with the 0\farcs5 slit was degraded due to problems within
the instrument. Soon after our observing run, CSHELL was serviced to
restore its spectral resolution to its nominal value. Our observations
achieve an actual resolution of $R \sim 24,000$, corresponding to a
FWHM of $\sim 4.5$ pixels on the $256 \times 256$ InSb array detector. 
A continuously variable filter (CVF) isolated individual orders of the
echelle grating, and a $\sim$ 0.0057 \micron\ portion of the spectrum 
was recorded in each of 3 wavelength settings. The first two settings 
(2.2218 and 2.2291 $\mu$m) each contain 2 magnetically sensitive 
\ion{Ti}{1} lines. The third setting (2.3125 $\mu$m) contains 9 
strong, magnetically insensitive CO lines. Each star was observed 
at two positions along the slit separated by 10\arcsec. Custom IDL
software described in Paper I was used to reduce the K-band
spectra. The reduction includes removal of the detector bias, dark
current, and night sky emission. Flat-fielding, cosmic ray removal,
and optimal extraction of the stellar spectrum are also included.
Wavelength calibration for each setting is based on a 3rd order
polynomial fitted to $n\lambda$ for several lamp emission lines
observed by changing the CVF while keeping the grating position
fixed. Dividing by the order number, $n$, for each setting yields 
the actual wavelength scale for the setting in question. In each
wavelength setting, we also observed a rapidly rotating early star to 
serve as a telluric standard.  All spectra shown here have been corrected
for telluric absorption.  The IR observations are summarized 
in Table \ref{obs}.

\section{Analysis and Results}

The analysis technique we employ here follows that in Papers I-III where
additional details can be found.
We model the profiles of four \ion{Ti}{1} absorption lines in the 
K band and measure the mean magnetic field (averaged over the 
entire stellar surface) for the five TWA stars. Diagnostics in the
IR are generally more sensitive to Zeeman broadening than lines in the 
optical thanks to the $\lambda^2$ dependence of magnetic splitting 
(see eq.[1]) compared with the $\lambda^1$ dependence of Doppler 
broadening mechanisms. Using estimates of the key stellar parameters 
of our objects, namely, \Teff, \logg, [M/H] and \vsini, we first 
construct model atmospheres and synthesize the \ion{Ti}{1} line 
profiles without the presence of magnetic fields. The observed 
magnetically insensitive CO lines are also synthesized to show that 
instrumental and rotational broadening are properly taken into 
account and the non-magnetic spectrum is accurately predicted. There 
are no free parameters involved in computing the non-magnetic line
profiles. The excess broadening seen in the \ion{Ti}{1} lines 
relative to the non-magnetic model is interpreted as Zeeman 
broadening. Fitting the magnetically broadened \ion{Ti}{1} line 
profiles then yields the average magnetic field strength on the 
stellar surface.

Paper III presented extensive tests of this analysis technique
to quantify possible systematic errors resulting from inaccurate
stellar parameters. The Monte Carlo analysis showed that a 200 K
error in \Teff\ or 0.5 dex error in \logg\ introduces only about 
$10\%$ or less systematic error in the derived mean field 
measurements. This is because we are modeling actual Zeeman 
broadening instead of some other effect such as a change in 
equivalent width which is particularly sensitive to errors in
\Teff\ and \logg. Thus, instead of performing a detailed spectral 
analysis to determine the stellar parameters as in our previous 
studies (e.g., Papers I-III), we adopt the most accurate values 
available in the literature for our targets. Values for these 
non-magnetic stellar parameters are given in Table \ref{results} 
and justified in subsequent subsections for each star.

The CO lines near $2.3\ \mu$m serve as an excellent verification of 
our chosen stellar parameters. For each of our targets, we 
interpolate a grid of the ``next generation'' (NextGen) atmospheres 
of \citet{allard95} to the specific values of \Teff, \logg\ and 
[M/H] and construct a model atmosphere appropriate for the star. We 
then synthesize the CO line profiles and convolve them with a 
Gaussian corresponding to the resolution of the observed spectrum. 
The CO line data are taken from \citet{goorvitch1994} for the CO 
{\it X} $\Sigma^+$ lines. No stellar parameter fitting is involved 
with the synthesis of the CO lines. Continuum normalization is 
the only free parameter used to match the observed and synthetic 
spectra. The shape, especially the widths, of the synthetic line 
profiles is solely determined by the stellar parameters. A good 
match between the widths of the synthetic CO lines and that 
of the observations, as seen in the bottom panels of Figures
1 and $3-6$, shows that the dominant non-magnetic broadening 
mechanisms are properly taken into account.

A polarized radiative transfer code \citep{piskunov1999}
is utilized to compute the line profiles for both the CO and
\ion{Ti}{1} lines, assuming 
a radial magnetic field geometry in the photosphere. To construct 
our models, we treat the stellar surface as if it is divided
into a number of regions, 
where one region is field-free and the other regions are each covered 
by a single magnetic field strength. The fractional size of each region 
is specified by a filling factor and the total of the filling factors 
must be unity. These regions are assumed to be spatially well mixed 
over the surface of the star. We compute the spectrum for each region, 
and the spectra of all regions are multiplied by their corresponding 
filling factors and summed up to form the final model spectrum for 
a specific set of field strengths and filling factors. We use 
the nonlinear least-squares technique of Marquardt 
\citep{bevington1992} to solve for the best-fit combination of field 
strengths and filling factors as described below.

For each target, we fit three different models. The first model 
(hereafter, M1) allows a single magnetic field strength ($B$) to 
cover a fraction of the stellar surface ($f$), and the rest of the 
surface is field-free. The mean stellar magnetic field, $\bar{B}$, 
is just $Bf$. The free parameters of this model are $B$ and $f$. 
The second model (hereafter, M2) divides the surface of the
star into three regions: two regions covered with magnetic fields of
different strengths ($B_1$, $B_2$) and filling factors ($f_1$, $f_2$), 
and one region with no magnetic field. The mean magnetic field 
on the stellar surface for M2 is $\bar B = B_1f_1 + B_2f_2$. The free
parameters of this model are $B_1$, $f_1$, $B_2$ and $f_2$. The third
model (hereafter, M3) allows three magnetic regions on 
the stellar surface to have specific field strengths of 2 kG, 4 kG and 
6 kG, respectively. The rest of the surface is field-free. Only the 
filling factors are solved for in this model. The mean field is then 
$\bar B = \displaystyle\sum_{i=1}^3 B_if_i$.


As discussed in Paper I-III, the M1 model provides a much better fit
to the observations than 
models with no magnetic broadening, but the model M1 profile is not 
as smooth and broad as the observed line profile, and is incapable 
of supplying good fits for the cores and the wings of the lines 
simultaneously. The magnetic structure on the surface of TTSs 
is unlikely to be so simple, and this is the motivation to add 
more magnetic components in model M2 and M3, which provides improved 
smoother fits to the observed line profiles. Due to lower spectral
resolution ($\sim 24,000$) than that in Paper I-III ($\sim 36,000$),
the need for more magnetic components is less urgent.  For example,
likely due to the lower spectral resolution achieved here, the data
do not clearly require a model with 3 magnetic components as in M3;
however, we fit such a 
model to be consistent with previous work in order to avoid any 
systematic differences when comparing the field measurements of TWA 
stars with previous results. 
Below, we discuss the results for each of our targets in detail.



\subsection{TWA 5A}

TWA 5 is among the five original TWA members reported by
\citet{kastner1997}. \citet{webb1999} discovered a M8.5 brown 
dwarf companion, TWA 5B, $\sim 2^{\prime\prime}$ away from the 
much brighter TWA 5A (M1.5). \citet{macintosh2001}, using 
adaptive optics (AO) imaging, found that TWA 5A itself is also 
a close binary system with a separation of $\sim0\farcs06$ at
that time. \citet{konopacky2007} monitored the TWA 5Aab system for 
five years. Their speckle and AO observations yield an orbital
solution with a semi-major axis of $0\farcs066 \pm 0\farcs005$ 
and a period of $5.94 \pm 0.09$ yrs.  These authors also find a K
band flux ratio of $\sim 1.25$, so we are most likely seeing 
substantial light from both stars in our spectra.

\begin{figure}[ht]
  \begin{center}
    \includegraphics[scale=0.60]{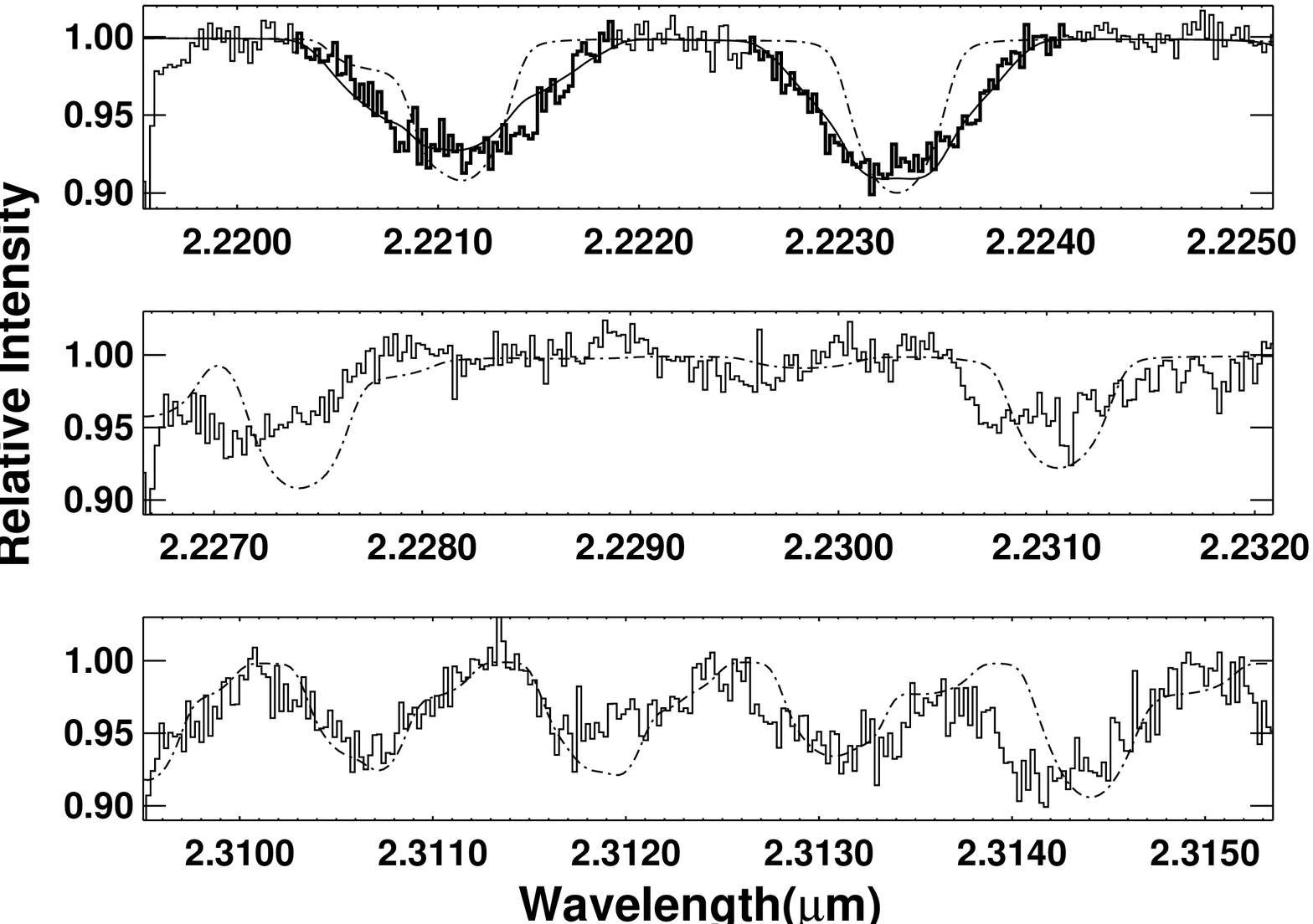}

   \caption{ Best fit of infrared spectra of TWA 5Aab system 
        using the two-magnetic-component model M2. 
	Top two panels: magnetically sensitive Ti I lines;
	Bottom panel: magnetically insensitive CO lines.
	(Histogram: data; dash-dotted line: fit without magnetic
	field; smooth line: fit with magnetic broadening.) }
	   \label{5plot}
	      \end{center}
	      \end{figure}

It is also suspected that there is possibly another spectroscopic
component in the TWA 5Aab pair, causing large radial velocity
variations \citep{torres2003}. We do notice peculiarities in our
spectra of TWA 5Aab system, which is shown in Figure \ref{5plot}. 
The three wavelength settings shown here were observed on 3
consecutive nights.
The two \ion{Ti}{1} line profiles in the $2.2300\ \mu$m setting 
are very asymmetric, and some of the CO lines show large wavelength 
shifts or asymmetries as well. With a period of about 6 years, 
the line profiles of the binary system obtained on consecutive 
nights should not vary so dramatically. This drastic change may 
be caused by a third component in the TWA 5Aab system in a close
orbit with one of the two known members.


We do not notice obvious asymmetry in the two \ion{Ti}{1} line 
profiles in the $2.2228\ \mu$m setting, so these two lines 
are used for magnetic field measurement. 
Following \citet{weintraub2000}, we adopt an effective 
temperature of 3700 K and derive a \logg\ of 3.7 from their absolute 
magnitude and stellar mass estimates, which assume a distance of 
$55 \pm 9$ pc \citep{webb1999}. A \vsini\ of 38 \kms\ is adopted
from \citet{torres2003}. The measured mean magnetic field values of 
TWA 5A along with that of the other four TWA stars, using the three 
different models, are listed in Table \ref{results}.  The reduced 
$\chi^2$ values for TWA 5A given in Table \ref{results} are calculated 
with all four \ion{Ti}{1} lines included so that it is consistent with other stars.
Note that only the two lines in the $2.2228\ \mu$m region are fitted and
the reduced $\chi^2$ values calculated with these two lines only are 
1.06, 1.10 and 1.28 for models M1-M3, repectively.
The observed and the best fit model 
spectra of TWA 5A are shown in Figure \ref{5plot}.  Given the
uncertainties associated with the number of stars in this system,
we report our field estimate here (which is likely some sort
of average of all the stars present) but do not use measurements
of this system in any subsequent analysis.


\subsection{TWA 7}

TWA 7 was discovered as a TWA member by \citet{webb1999}.
\citet{neuhauser2000} detected a possible extra-solar planet
candidate about 2\farcs5 away from TWA 7 using ground-based direct
imaging, but they concluded that this object is more likely an 
unrelated background source. This interpretation was later 
confirmed by the coronagraphic observations of \citet{lowrance2005}.
An inclination of $28^\circ$ is estimated for TWA 7.

\begin{figure}[ht] 
  \begin{center}
    \includegraphics[scale=0.75]{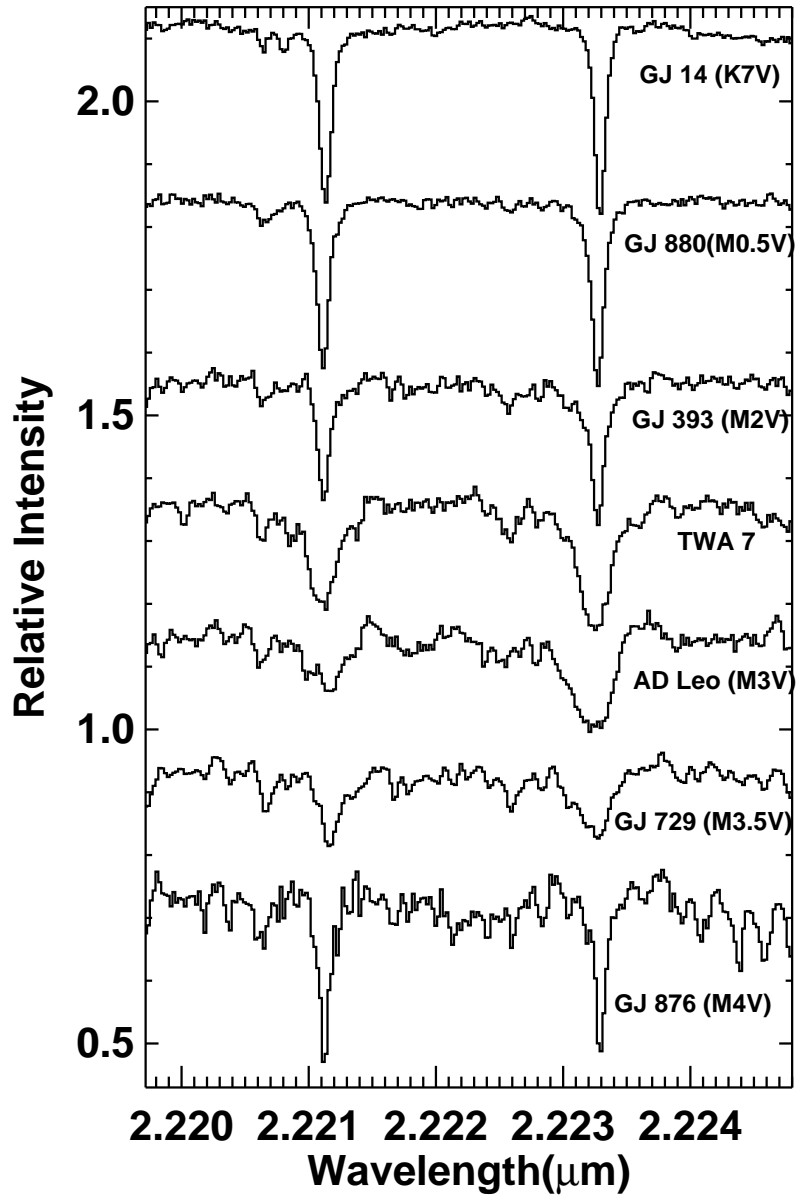}

       \caption{Spectra of TWA 7 along with several spectral type 
                standards in the $2.2228\ \mu$m setting. The spectra
		are all vertically offset for better display. Note that
		both AD Leo and GJ 729 are known to have strong 
		magnetic fields \citep{cmj2000}. See also earlier
		measurements for AD Leo by \citet{saar1985}, and 
		GJ 729 by \citet{cmj1996}.} 
	   \label{sptype}
	      \end{center}
    \end{figure}

\begin{figure}[ht] 
  \begin{center}
    \includegraphics[scale=0.60]{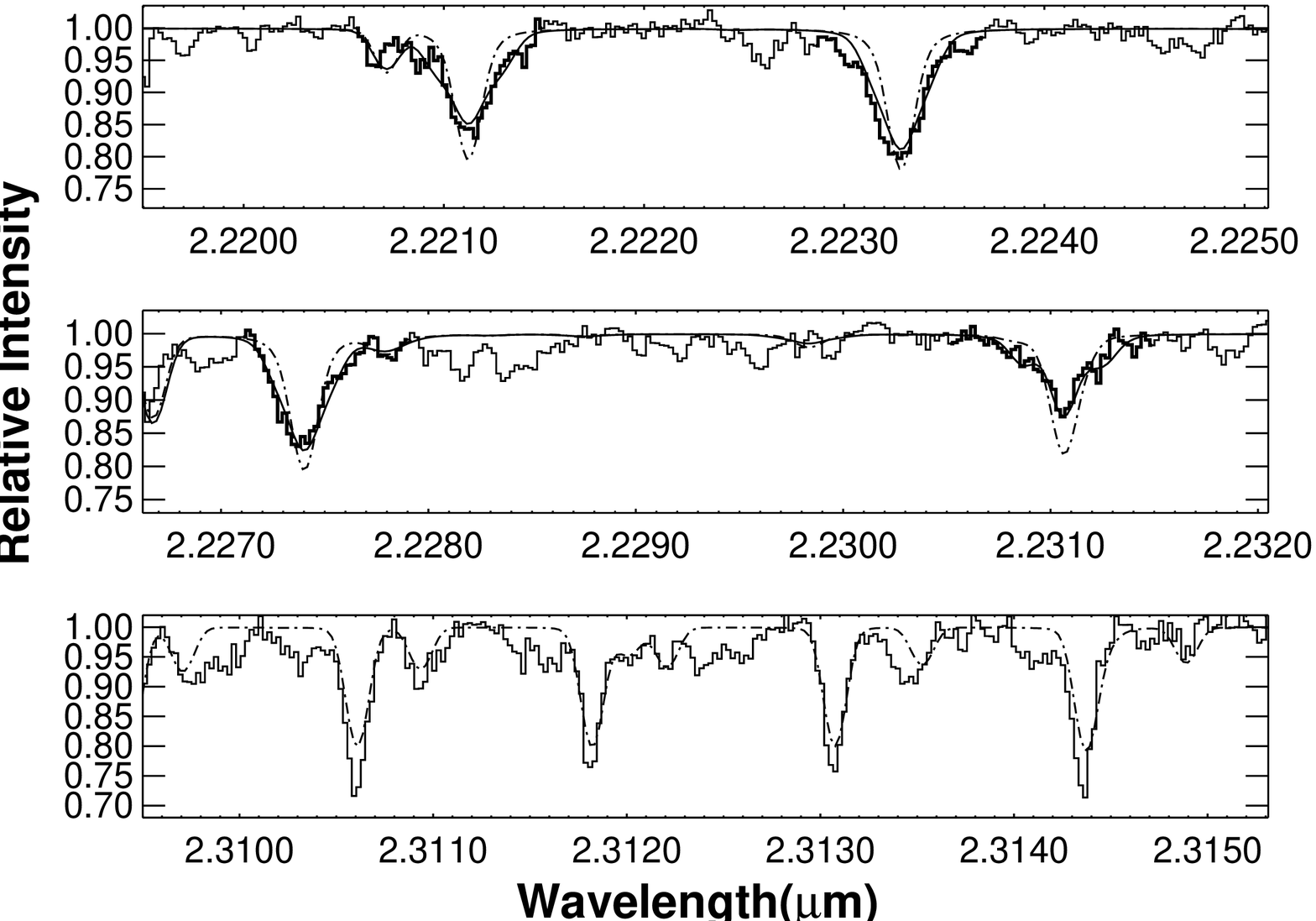}

   \caption{ Best fit of infrared spectra of TWA 7  using the 
        two-magnetic-component model M2. 
	Top two panels: magnetically sensitive Ti I lines;
	Bottom panel: magnetically insensitive CO lines.
	(Histogram: data; dash-dotted line: fit without magnetic
	field; smooth line: fit with magnetic broadening.) }

	   \label{7plot}
	      \end{center}
	      \end{figure}

\citet{webb1999} determined a spectral type of M1 for TWA 7 by 
comparing their low resolution optical spectra with spectral 
standards from \citet{montes1998}. In our high resolution IR data, 
we notice that some molecular lines of TWA $7$ in the K band are 
more prominent than that indicated by a M1 spectral type. Shown 
in Figure \ref{sptype}, we compare the spectrum of TWA $7$ in the 
$2.2228\ \mu$m setting with that of several stars of different 
spectral types. From this TWA 7 appears to be of later spectral 
type than M1, probably closer to M3. Though we do not synthsize 
molecular lines in this wavelength setting, the \ion{Ti}{1} lines 
are not seriously blended with these features.  The molecular 
lines, possibly CN, do not substantially affect our magnetic 
field measurement. In our analysis we adopt a \Teff\ of 3300 K 
for TWA 7, appropriate for an M3 star, and a \logg\ $= 3.85$ is 
derived by adopting the bolometric luminosity and stellar mass 
estimates of \citet{neuhauser2000}. The conversion of spectral 
type to effective temperature in this paper is based on 
\citet{johnson1966}. Interestingly, Johnson’s spectral type to \Teff\
conversions closely parallel those derived for young stars by 
\citet{luhman2003}, from G through mid-M spectral types.
We use a \vsini\ $= 9.0$ \kms, which is 
slightly larger than previous measured value of $6.1 \pm 0.5$ 
\kms\ from \citet{neuhauser2000}. Our choices of \vsini\ are 
based initially on literature values and in some cases they 
are adjusted by a small amount to better match the width of 
the observed nonmagnetic CO lines. As mentioned above, our 
analysis technique suffers $\sim 10\%$ or less systematic error 
for a 200 K error in \Teff\ or 0.5 dex error in \logg, so the 
measured field values of TWA 7 are not subject to much change, 
whether its spectral type is M1 or M3. The best fit for TWA 7 is 
shown in Figure \ref{7plot}.


\subsection{TWA 8A}

\begin{figure}[ht]
  \begin{center}
    \includegraphics[scale=0.60]{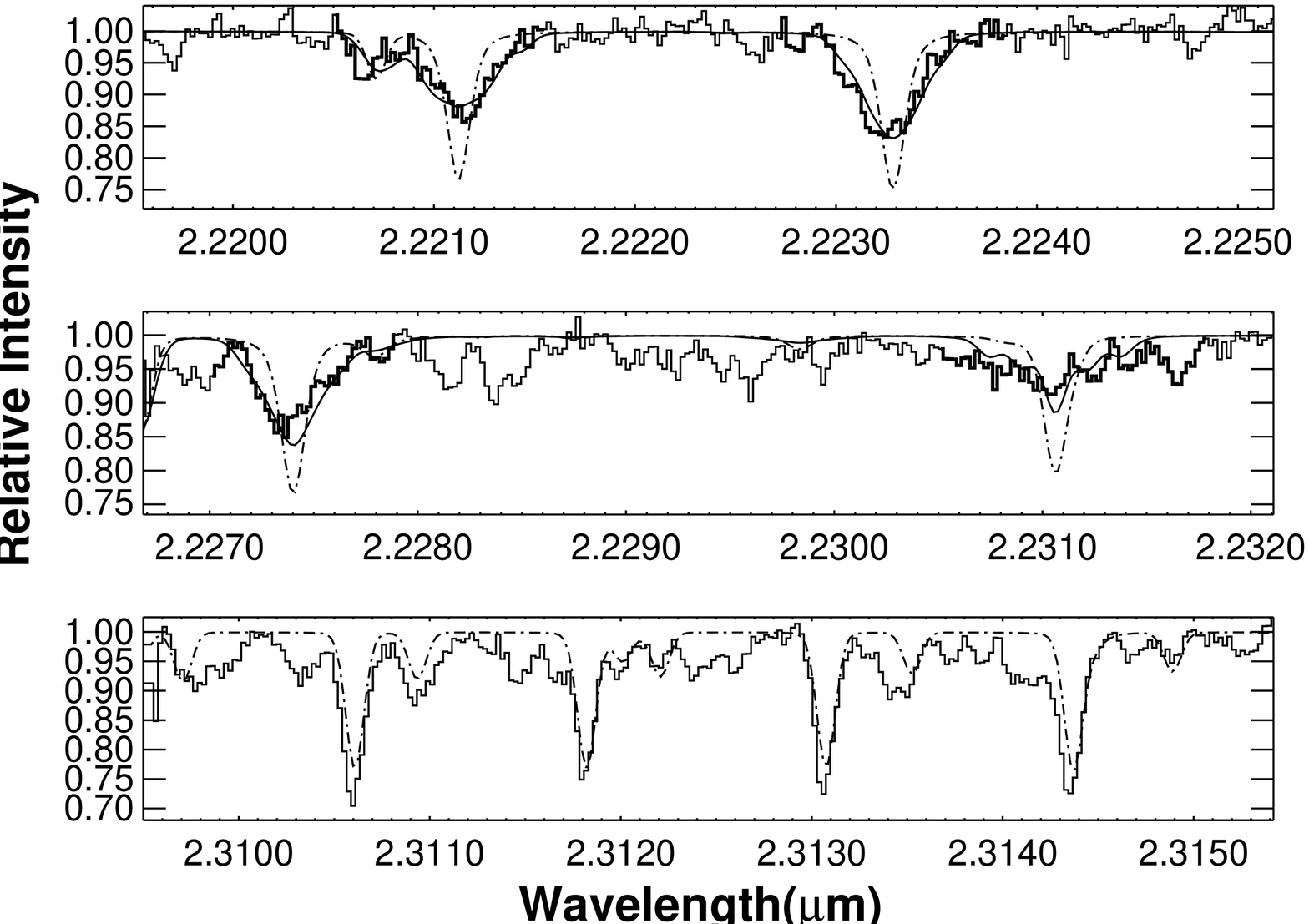}

   \caption{ Best fit of infrared spectra of TWA 8A using the 
        two-magnetic-component model M2. 
	Top two panels: magnetically sensitive Ti I lines;
	Bottom panel: magnetically insensitive CO lines.
	(Histogram: data; dash-dotted line: fit without magnetic
	field; smooth line: fit with magnetic broadening.) }

	   \label{8aplot}
	      \end{center}
	      \end{figure}

TWA 8A was also identified by \citet{webb1999} as an association 
member, and is classified as a M2 star. Its fainter companion 
TWA 8B is about $13 \arcsec$ to the south of the primary. The 
membership of the companion in TWA is based primarily on its similar lithium 
abundance and radial velocity compared to other members of the 
association. Trigonometric parallax and proper motion measurements 
are needed for further confirmation.  \citet{white2004} derive 
a spectral type of M3 ($\pm 0.5$) for TWA 8A, by matching line 
and molecular band ratios to those of dwarf spectral standards 
which are rotationally broadened.  These authors also determined 
an upper limit of 5.5\ \kms\ for the \vsini\ of TWA 8A. For our 
analysis, we use a \Teff\ of 3400 K, \logg\ of 4.0, and \vsini\ 
of 4.0 \kms. The inclination of TWA 8A is estimated to be $16^\circ$,
so it is seen close to pole-on. The best fit for the \ion{Ti}{1} lines as well as 
the predicted CO spectrum of TWA 8A are shown compared with 
the observed spectra in Figure \ref{8aplot}.

\subsection{TWA 9A \& 9B}

TWA 9 (CD $-36^{\circ}7429$) was found by \citet{jensen1998} to be in 
the physical vicinity of TW Hya. It is a binary system with a 
separation of $6 \arcsec$. The \textit{Hipparcos} 
satellite \citep{esa1997} measured a distance of $50.3 \pm 6.0$ pc 
for TWA 9 system. In the literature, TWA 9A is classified either as K5 
\citep{webb1999} or K7 ($\pm\ 1 $) \citep{white2004}. Previously reported 
\vsini\ values are $7 \pm 3\ \kms $ \citep{reid2003} and $11 \pm 1\ \kms\ 
$ \citep{white2004} for TWA 9A. We adopt \Teff\ = 4000 K and \logg\ = 4.5, 
and use \vsini\ = 9 \kms. The inclination angle of TWA 9A is inferred to
be $90^\circ$.

\begin{figure}[ht]
  \begin{center}
    \includegraphics[scale=0.60]{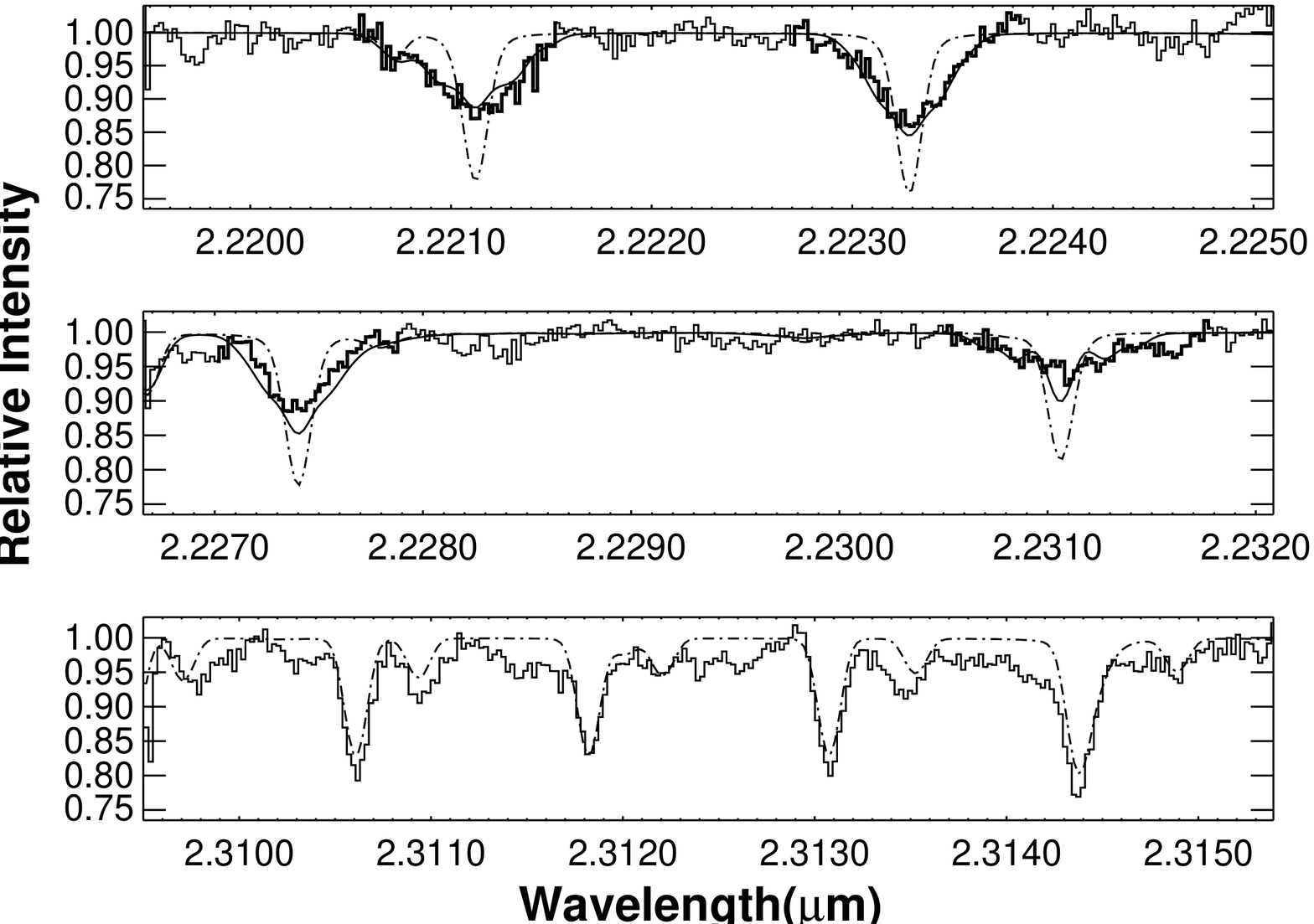}

   \caption{ Best fit of infrared spectra of TWA 9A using the 
        two-magnetic-component model M2. 
	Top two panels: magnetically sensitive Ti I lines;
	Bottom panel: magnetically insensitive CO lines.
	(Histogram: data; dash-dotted line: fit without magnetic
	field; smooth line: fit with magnetic broadening.) }
	   \label{a9plot}
	      \end{center}
	      \end{figure}

\begin{figure}[ht]
  \begin{center}
    \includegraphics[scale=0.60]{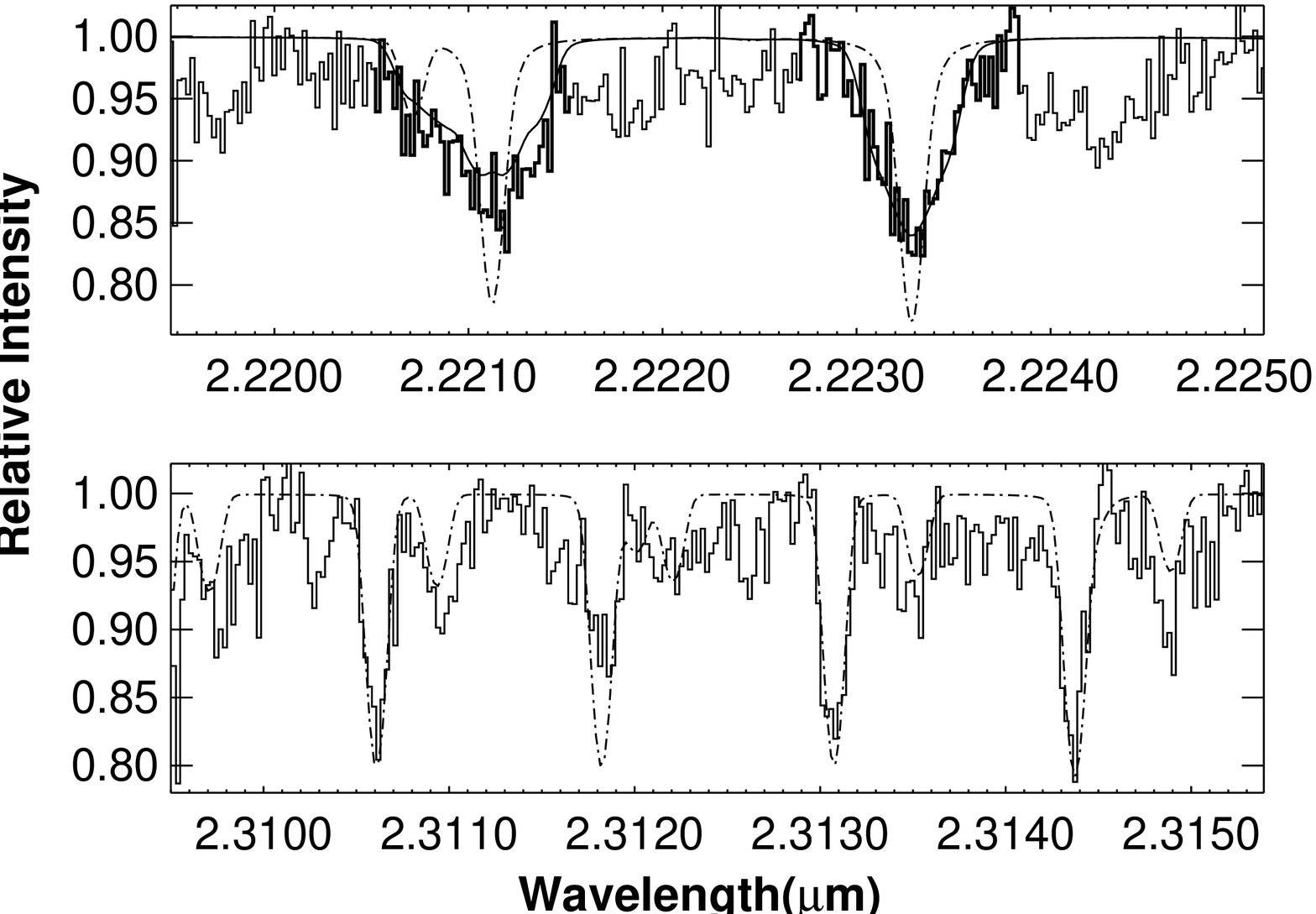}

   \caption{ Best fit of infrared spectra of TWA 9B using the 
        two-magnetic-component model M2. 
	Top panel: magnetically sensitive Ti I lines;
	Bottom panel: magnetically insensitive CO lines.
	(Histogram: data; dash-dotted line: fit without magnetic
	field; smooth line: fit with magnetic broadening.) }
	   \label{b9plot}
	      \end{center}
	      \end{figure}

For TWA 9B, reported spectral types include M1 \citep{webb1999} and 
M3.5 ($\pm\ 0.5$) \citep{white2004}, and \vsini\ values of $4 \pm 3\ \kms$ 
\citep{reid2003} as well as $9 \pm 1\ \kms$ \citep{white2004} have been
published. We use \Teff\ = 3400 K, \logg\ = 4.0, and \vsini\ = 10 \kms. 
The inclination angle is estimated to be $60^\circ$.
Due to the very poor signal-to-noise ratio we obtain on this star, the 
spectrum in the $2.2300\ \mu$m setting is excluded from our analysis.  
The best fit spectra along with observed spectra of TWA 9A and TWA 9B are 
shown in Figure \ref{a9plot} and \ref{b9plot}, respectively.

\section{Discussion}

By fitting the Zeeman broadening in the magnetically sensitive 
\ion{Ti}{1} lines, we measure the average surface magnetic field 
of five TWA stars.  Three different models were used to fit the
line profiles and measure the magnetic field strength with generally
good agreement in the results between the three.  Model M1 assumes
only 1 magnetic component to the stellar atmosphere.  This model
provides a good match to TWA 5A in the $2.2228 \mu$m region (but not
the $2.2300 \mu$m region).
This is likely due to the fact that TWA 5A apparently 
has a relatively large $v$sin$i$ so that rotational broadening,
in addition to the magnetic broadening, strongly affects the line 
profile shapes. Because our magnetic models for TWA 5A do not reproduce
line profiles in the $2.2300 \mu$m region, our derived magnetic properties 
for TWA 5A may have large systematic errors. Generally, mean field 
strength increases as more components are added to the model, which suggests
that the observed profiles have wings that are too broad to be fit by a 
single component. For the other stars, the lowest reduced $\chi^2$
typically occurs for model M2, though M3 is favored in
one case.  However, there is generally quite good agreement bewteen the
mean field values recovered using these two models with the difference
bewteen them averaging 15\%.  For consistency with previous studies,
we use the results from M3 when comparing to results found by
\citet{cmj2007}.

Theoretically, equipartition arguments for the Sun and cool stars suggest 
that the magnetic field strength should scale with the gas pressure 
in the surrounding non-magnetic photosphere which sets a limit for 
the maximum field strength allowed on the star
\citep[e.g.][]{spruit1979,safier1999}.
\citet{rajaguru2002} found that in the regime 
where convectively stable flux tubes exist, equipartition pressure 
allows a maximum magnetic field strength of $\sim 1300$ G for values 
of \Teff\ and \logg\ appropriate for most of our TWA stars. We calculate 
equipartition field strengths for the TWA stars by balancing 
magnetic pressure, $B_{eq}^2/(8\pi)$, with the gas pressure in 
surrounding unmagnetized photosphere, $P_{g}$. We use the gas 
pressure at the level in the atmosphere where the local temperature 
is equal to the effective temperature in the model atmospheres 
used for our stars. This should be an upper limit for gas 
pressure because this level is approximately where the continuum 
forms, while the \ion{Ti}{1} lines form over a wide range of depths 
above this level in the atmosphere. The equipartition field 
strengths, $B_{eq}={(8\pi P_{g})}^{1/2}$, of the TWA stars are 
listed in the last column of Table \ref{results}. These values are 
in good agreement with the estimate from \citet{rajaguru2002}. 
Compared with the measured mean field values, they are all smaller 
by a factor of at least 1.38 and more typically $\sim 2.0$ (for model M2), 
suggesting that magnetic pressure 
dominates over gas pressure in the photospheres of the TWA stars observed 
here. This in turn implies complete coverage of magnetic fields over 
surface of these stars. Magnetic regions with excess pressure would expand,
while field free regions contract
into the lower pressure nonmagnetic regions until equipartition is 
established at a lower field strength or until the entire star is covered 
with magnetic fields. As discussed in Papers I and II, 
our models do not necessarily require field-free regions. The 
results are essentially the same with or without a 0 kG field 
component. As an example, we fit the observed spectra of TWA 8A 
with a model which allows on the stellar surface no field-free 
region and four magnetic regions with specific field strengths 
of 1 kG, 3 kG, 5 kG and 7 kG. The measured magnetic field value 
using this model is 3.4 kG with a reduced $\chi^2$ of 1.22, while 
model M3 yields 3.3 kG with a reduced $\chi^2$ of 1.25
(see Table \ref{results}). With a field-free region 
included in our current analysis, all the TWA stars have at 
least $76\%$, in most cases $80\% - 86\%$, of their surfaces 
covered with magnetic fields. The surfaces of TTSs are quite 
possibly fully covered by strong magnetic fields.

\begin{figure}[ht] 
  \begin{center}
    \includegraphics[scale=0.60]{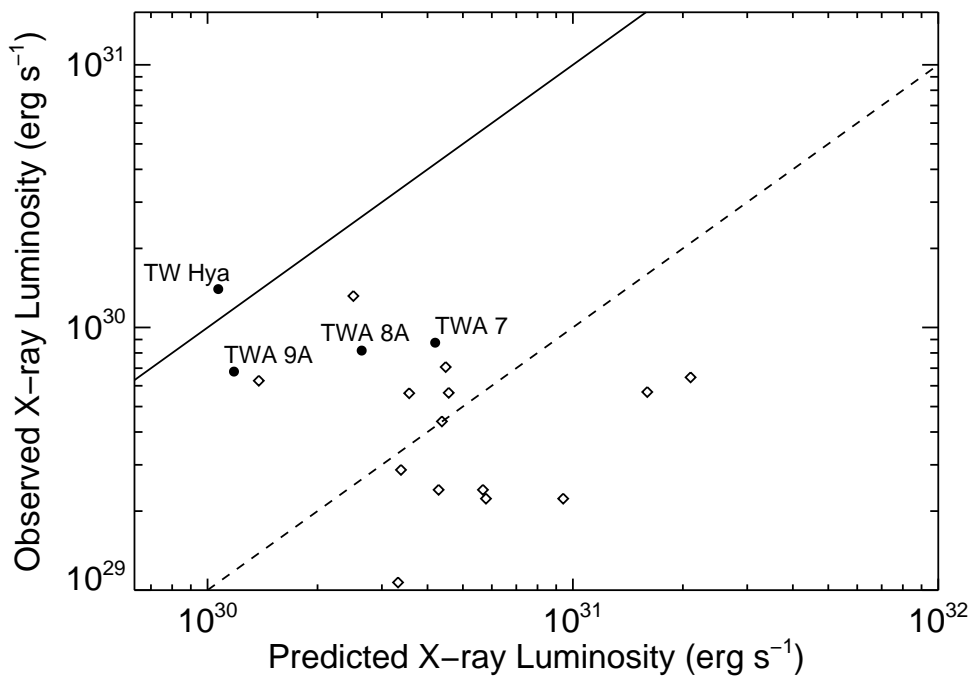}

   \caption{ Observed X-ray luminosity plotted against the predicted 
         X-ray luminosity for the TWA stars. The solid line is the 
	 line of equality, and the dashed line is where the predicted 
	 value is 10 times the observed one. The filled circles represent 
         the TWA stars from this paper, with the observed X-ray luminosity 
         derived from \citet{webb1999}. The hollow diamonds represent 
         the Taurus \& Auriga stars from \citet{cmj2007}. }

	   \label{xray}
	      \end{center}
	      \end{figure}

\citet{pevtsov2003} examine the X-ray luminosities and the total 
unsigned magnetic flux of active stars as well as regions of 
different activity level on the Sun. They find a nearly linear 
correlation over many decades of variation in the two quantities. 
We calculate the magnetic flux of the four TWA stars (excluding
TWA 5A since it is an unresolved multiple system), using our magnetic field 
measurements and the stellar radii estimated from bolometric 
luminosities given in \citet{webb1999} and the effective 
temperatures of our stars. We utilize the X-ray 
luminosity-magnetic flux relationship of \citet{pevtsov2003} and 
calculate the predicted X-ray luminosity. For comparison, we 
use the X-ray to bolometric luminosity ratio, $L_x/L_{bol}$, 
reported by \citet{webb1999} to get the value of $L_x$ for each
star. The predicted and measured values of $L_x$ for the three 
TWA stars (TWA 7, TWA 8A and TWA 9A) presented in this paper 
and that of TW Hydrae taken from Paper III are plotted in Figure 
\ref{xray}. A measured value for TWA 9B is not available in the 
literature. As shown in Figure. \ref{xray}, the observed $L_x$
values of the four TWA stars are smaller than the predicted 
values by a factor of $2.7 \pm 0.7$ on average. This is similar 
to the findings of \citet{cmj2007} in a study of 14 CTTSs in 
Taurus and Auriga. The stars from \citet{cmj2007} are also 
plotted in Figure \ref{xray}, and the difference between the 
observed and the predicted $L_x$ values is a bit more obvious. 
The average ratio of the predicted to observed $L_x$ values is 
$17.7 \pm 3.4$ for the 14 Taurus and Auriga stars. 
At an age of about 10 Myrs, 
the TWA stars are substantially older than the stars in the 
Taurus-Auriga region ( $\sim$2 Myrs). As a result, the TWA 
stars are substantially closer to the main sequence and this 
may be the reason why their X-ray luminosities and magnetic 
flux values are closer to the relationship found by 
\citet{pevtsov2003} which was based primarily on main sequence 
stars. The X-ray emission on active stars such as TTSs is 
thought to mainly result from energy dissipation in magnetic 
fields during flares and small-scale flare-like activities 
\citep{guedel2003,arzner2007}. Magnetic fields are stressed due
to convective gas motions in the lower photosphere and below
\citep{foukal2004}, producing flares of various scales. 
For PMS stars that are fully covered by magnetic fields, when
the magnetic field strength is larger than the equipartition 
value, it may be more difficult for the gas, at least in the photosphere, 
to push around the fields and build up magnetic stress. This would be the case 
for both the Taurus stars and the TWA stars. The ratio of the
observed to equipartion magnetic field, $B_{obs}/B_{eq}$, 
for the Taurus stars on average is greater than that for the 
TWA stars, so the X-ray production might be more prohibited 
for the Taurus stars resulting in the observed $L_x$ being further away from 
the line of equality in Figure \ref{xray}. 
It will be interesting 
to see where TTSs in other age groups fit in this picture 
relating magnetic flux and X-ray emission.


The origin of magnetic fields on TTSs has sparked various theoretical 
speculations, but it is still far from clear how the strong fields on 
these stars are produced. Due to their youth, most TTSs are generally
thought to be fully convective, though model calculations have shown
that TTSs might have small radiative cores \citep{granzer2000}.
As a result, a solar-like interface dynamo 
\citep[e.g.,][]{durney1978,parker1993} is unlikely to be responsible 
for the fields on TTSs. One potential alternative mechanism is a 
distributed dynamo amplified by convective motions 
\citep[e.g.,][]{durney1993, dobler2006, chabrier2006}.  Such turbulent
dynamos are not well understood at this point, but exisiting models
seem to only produce field strengths equal to or lower than the
pressure equipartition values, and many such models also produce strong
fields which occupy only a small filling factor at the stellar surface
\citep[e.g.,][]{cattaneo1999, bercik2005}.  Both of these predicitons
appear not to be borne out by the observations.  In addition to a possible
dynamo origin, the magnetic fields of TTSs could also result from the 
primordial flux that survives the star formation process 
\citep[e.g.,][]{tayler1987, moss2003}. 
With the measurements from this paper and 
those in \citet{yang2005} and \citet{cmj2007}, we explore this issue 
by comparing the average magnetic field strength and magnetic flux of 
the stars in TWA and in Taurus-Auriga, two regions of different ages. 
The average magnetic field of the Taurus-Auriga stars is 
$2.20 \pm 0.16$ kG, which is smaller than that of the five TWA stars 
(TW Hya, TWA 7, TWA 8A, TWA 9A and TWA 9B), where 
$\bar B =  2.98 \pm 0.22$ kG. While the average 
field strength increases slightly from the age of Taurus to that of TWA,
we see a sharp decrease of the average 
magnetic flux, $4\pi {R_*}^2B$, from $(7.00 \pm 1.37) \times 10^{18}$ Wb 
for the Taurus-Auriga stars compared to $(2.61 \pm 0.69) \times 10^{18}$ 
Wb for the five stars in TWA. This change in the magnetic flux is 
generally consistent with the decay of primordial magnetic fields 
on TTSs as suggested by \citet{moss2003}. With only a limited number 
of samples on hand, the current evidence is far from sufficient. Further 
observations and measurements of young stars in other young clusters 
of different ages, e.g., the Orion Nebula Cluster and the $\beta$ 
Pictoris Moving Group, could shed more light on the issue of field
generation in young stars.


\acknowledgements

C.M.J.-K. and H. Y. would like to acknowledge partial support from the 
NASA Origins of Solar Systems program through grant numbers NAG5-13103 
and NNG06GD85G.  The authors extend special thanks to those of Hawaiian 
ancestry on whose sacred mountain we are privileged to be guests. This 
work made use of the SIMBAD reference database, the NASA Astrophysics 
Data System.

\bibliography{twaref}

\begin{thebibliography}{80}
\expandafter\ifx\csname natexlab\endcsname\relax\def\natexlab#1{#1}\fi

\bibitem[{{Allard} \& {Hauschildt}(1995)}]{allard95}
{Allard}, F., \& {Hauschildt}, P.~H. 1995, in The Bottom of the Main Sequence -
  and Beyond, Proceedings of the ESO Workshop Held in Garching, Germany, 10-12
  August 1994, edited by Christopher G. Tinney. Springer-Verlag Berlin
  Heidelberg New York. Also ESO Astrophysics Symposia, 1995., p.32, ed. C.~G.
  {Tinney}, 32--+

\bibitem[{{Appenzeller} \& {Mundt}(1989)}]{appenzeller1989}
{Appenzeller}, I., \& {Mundt}, R. 1989, \aapr, 1, 291

\bibitem[{{Arzner} {et~al.}(2007){Arzner}, {G{\"u}del}, {Briggs}, {Telleschi},
  \& {Audard}}]{arzner2007}
{Arzner}, K., {G{\"u}del}, M., {Briggs}, K., {Telleschi}, A., \& {Audard}, M.
  2007, \aap, 468, 477

\bibitem[{{Barrado Y Navascu{\'e}s}(2006)}]{navascues2006}
{Barrado Y Navascu{\'e}s}, D. 2006, \aap, 459, 511

\bibitem[{{Basri} \& {Bertout}(1993)}]{basribertout1993}
{Basri}, G., \& {Bertout}, C. 1993, in Protostars and Planets III, ed. E.~H.
  {Levy} \& J.~I. {Lunine}, 543--566

\bibitem[{{Basri} {et~al.}(1992){Basri}, {Marcy}, \& {Valenti}}]{basri1992}
{Basri}, G., {Marcy}, G.~W., \& {Valenti}, J.~A. 1992, \apj, 390, 622

\bibitem[{{Bercik} {et~al.}(2005){Bercik}, {Fisher}, {Johns-Krull}, \&
  {Abbett}}]{bercik2005}
{Bercik}, D.~J., {Fisher}, G.~H., {Johns-Krull}, C.~M., \& {Abbett}, W.~P.
  2005, \apj, 631, 529

\bibitem[{{Bertout}(1989)}]{bertout1989}
{Bertout}, C. 1989, \araa, 27, 351

\bibitem[{{Bevington} \& {Robinson}(1992)}]{bevington1992}
{Bevington}, P.~R., \& {Robinson}, D.~K. 1992, {Data reduction and error
  analysis for the physical sciences} (New York: McGraw-Hill, |c1992, 2nd ed.)

\bibitem[{{Bouvier} {et~al.}(2007){Bouvier}, {Alencar}, {Harries},
  {Johns-Krull}, \& {Romanova}}]{bouvier2007}
{Bouvier}, J., {Alencar}, S.~H.~P., {Harries}, T.~J., {Johns-Krull}, C.~M., \&
  {Romanova}, M.~M. 2007, in Protostars and Planets V, ed. B.~{Reipurth},
  D.~{Jewitt}, \& K.~{Keil}, 479--494

\bibitem[{{Camenzind}(1990)}]{camenzind1990}
{Camenzind}, M. 1990, in Reviews in Modern Astronomy, ed. G.~{Klare}, 234--265

\bibitem[{{Cattaneo}(1999)}]{cattaneo1999}
{Cattaneo}, F. 1999, \apjl, 515, L39

\bibitem[{{Chabrier} \& {K{\"u}ker}(2006)}]{chabrier2006}
{Chabrier}, G., \& {K{\"u}ker}, M. 2006, \aap, 446, 1027

\bibitem[{{Collier Cameron} \& {Campbell}(1993)}]{ccc1993}
{Collier Cameron}, A., \& {Campbell}, C.~G. 1993, \aap, 274, 309

\bibitem[{{Daou} {et~al.}(2006){Daou}, {Johns-Krull}, \& {Valenti}}]{daou2006}
{Daou}, A.~G., {Johns-Krull}, C.~M., \& {Valenti}, J.~A. 2006, \aj, 131, 520

\bibitem[{{Dobler} {et~al.}(2006){Dobler}, {Stix}, \&
  {Brandenburg}}]{dobler2006}
{Dobler}, W., {Stix}, M., \& {Brandenburg}, A. 2006, \apj, 638, 336

\bibitem[{{Donati} {et~al.}(2007){Donati}, {Jardine}, {Gregory}, {Petit},
  {Bouvier}, {Dougados}, {M{\'e}nard}, {Cameron}, {Harries}, {Jeffers}, \&
  {Paletou}}]{donati2007}
{Donati}, J.-F., {Jardine}, M.~M., {Gregory}, S.~G., {Petit}, P., {Bouvier},
  J., {Dougados}, C., {M{\'e}nard}, F., {Cameron}, A.~C., {Harries}, T.~J.,
  {Jeffers}, S.~V., \& {Paletou}, F. 2007, \mnras, 380, 1297

\bibitem[{{Donati} {et~al.}(2008){Donati}, {Jardine}, {Gregory}, {Petit},
  {Paletou}, {Bouvier}, {Dougados}, {M{\'e}nard}, {Cameron}, {Harries},
  {Hussain}, {Unruh}, {Morin}, {Marsden}, {Manset}, {Auri{\`e}re}, {Catala}, \&
  {Alecian}}]{donati2008}
{Donati}, J.-F., {Jardine}, M.~M., {Gregory}, S.~G., {Petit}, P., {Paletou},
  F., {Bouvier}, J., {Dougados}, C., {M{\'e}nard}, F., {Cameron}, A.~C.,
  {Harries}, T.~J., {Hussain}, G.~A.~J., {Unruh}, Y., {Morin}, J., {Marsden},
  S.~C., {Manset}, N., {Auri{\`e}re}, M., {Catala}, C., \& {Alecian}, E. 2008,
  \mnras, 386, 1234

\bibitem[{{Durney} {et~al.}(1993){Durney}, {De Young}, \&
  {Roxburgh}}]{durney1993}
{Durney}, B.~R., {De Young}, D.~S., \& {Roxburgh}, I.~W. 1993, \solphys, 145,
  207

\bibitem[{{Durney} \& {Latour}(1978)}]{durney1978}
{Durney}, B.~R., \& {Latour}, J. 1978, Geophysical and Astrophysical Fluid
  Dynamics, 9, 241

\bibitem[{{Edwards} {et~al.}(1994){Edwards}, {Hartigan}, {Ghandour}, \&
  {Andrulis}}]{edwards1994}
{Edwards}, S., {Hartigan}, P., {Ghandour}, L., \& {Andrulis}, C. 1994, \aj,
  108, 1056

\bibitem[{{Eisner} {et~al.}(2005){Eisner}, {Hillenbrand}, {White}, {Akeson}, \&
  {Sargent}}]{eisner2005}
{Eisner}, J.~A., {Hillenbrand}, L.~A., {White}, R.~J., {Akeson}, R.~L., \&
  {Sargent}, A.~I. 2005, \apj, 623, 952

\bibitem[{{ESA}(1997)}]{esa1997}
{ESA}. 1997, VizieR Online Data Catalog, 1239, 0

\bibitem[{{Foukal}(2004)}]{foukal2004}
{Foukal}, P.~V. 2004, {Solar Astrophysics, 2nd, Revised Edition} (Solar
  Astrophysics, 2nd, Revised Edition, by Peter V.~Foukal, pp.~480.~ISBN
  3-527-40374-4.~Wiley-VCH , April 2004.)

\bibitem[{{Goorvitch}(1994)}]{goorvitch1994}
{Goorvitch}, D. 1994, \apjs, 95, 535

\bibitem[{{Granzer} {et~al.}(2000){Granzer}, {Sch{\"u}ssler}, {Caligari}, \&
  {Strassmeier}}]{granzer2000}
{Granzer}, T., {Sch{\"u}ssler}, M., {Caligari}, P., \& {Strassmeier}, K.~G.
  2000, \aap, 355, 1087

\bibitem[{{Greene} {et~al.}(1993){Greene}, {Tokunaga}, {Toomey}, \&
  {Carr}}]{greene93}
{Greene}, T.~P., {Tokunaga}, A.~T., {Toomey}, D.~W., \& {Carr}, J.~B. 1993, in
  Proc. SPIE Vol. 1946, p. 313-324, Infrared Detectors and Instrumentation,
  Albert M. Fowler; Ed., ed. A.~M. {Fowler}, 313--324

\bibitem[{{G{\"u}del} {et~al.}(2003){G{\"u}del}, {Audard}, {Kashyap}, {Drake},
  \& {Guinan}}]{guedel2003}
{G{\"u}del}, M., {Audard}, M., {Kashyap}, V.~L., {Drake}, J.~J., \& {Guinan},
  E.~F. 2003, \apj, 582, 423

\bibitem[{{Guenther} {et~al.}(1999){Guenther}, {Lehmann}, {Emerson}, \&
  {Staude}}]{guenther1999}
{Guenther}, E.~W., {Lehmann}, H., {Emerson}, J.~P., \& {Staude}, J. 1999, \aap,
  341, 768

\bibitem[{{Hartmann} {et~al.}(1994){Hartmann}, {Hewett}, \&
  {Calvet}}]{hartmann1994}
{Hartmann}, L., {Hewett}, R., \& {Calvet}, N. 1994, \apj, 426, 669

\bibitem[{{Huelamo} {et~al.}(2008){Huelamo}, {Figueira}, {Bonfils}, {Santos},
  {Pepe}, {Guillon}, {Azevedo}, {Barman}, {Fernandez}, {di Folco}, {Guenther},
  {Lovis}, {Melo}, {Queloz}, \& {Udry}}]{huelamo2008}
{Huelamo}, N., {Figueira}, P., {Bonfils}, X., {Santos}, N.~C., {Pepe}, F.,
  {Guillon}, M., {Azevedo}, R., {Barman}, T., {Fernandez}, M., {di Folco}, E.,
  {Guenther}, E.~W., {Lovis}, C., {Melo}, C.~H.~F., {Queloz}, D., \& {Udry}, S.
  2008, ArXiv e-prints, 808

\bibitem[{{Jensen} {et~al.}(1998){Jensen}, {Cohen}, \&
  {Neuh{\"a}user}}]{jensen1998}
{Jensen}, E.~L.~N., {Cohen}, D.~H., \& {Neuh{\"a}user}, R. 1998, \aj, 116, 414

\bibitem[{{Johns-Krull}(2007)}]{cmj2007}
{Johns-Krull}, C.~M. 2007, \apj, 664, 975

\bibitem[{{Johns-Krull} \& {Valenti}(1996)}]{cmj1996}
{Johns-Krull}, C.~M., \& {Valenti}, J.~A. 1996, \apjl, 459, L95+

\bibitem[{{Johns-Krull} \& {Valenti}(2000)}]{cmj2000}
{Johns-Krull}, C.~M., \& {Valenti}, J.~A. 2000, in Astronomical Society of the
  Pacific Conference Series, Vol. 198, Stellar Clusters and Associations:
  Convection, Rotation, and Dynamos, ed. R.~{Pallavicini}, G.~{Micela}, \&
  S.~{Sciortino}, 371--+

\bibitem[{{Johns-Krull} {et~al.}(1999a){Johns-Krull}, {Valenti}, {Hatzes}, \&
  {Kanaan}}]{cmj1999a}
{Johns-Krull}, C.~M., {Valenti}, J.~A., {Hatzes}, A.~P., \& {Kanaan}, A. 1999a,
  \apjl, 510, L41

\bibitem[{{Johns-Krull} {et~al.}(1999b){Johns-Krull}, {Valenti}, \&
  {Koresko}}]{cmj1999b}
{Johns-Krull}, C.~M., {Valenti}, J.~A., \& {Koresko}, C. 1999b, \apj, 516, 900

\bibitem[{{Johns-Krull} {et~al.}(2004){Johns-Krull}, {Valenti}, \&
  {Saar}}]{cmj2004}
{Johns-Krull}, C.~M., {Valenti}, J.~A., \& {Saar}, S.~H. 2004, \apj, 617, 1204

\bibitem[{{Johnson}(1966)}]{johnson1966}
{Johnson}, H.~L. 1966, \araa, 4, 193

\bibitem[{{Kastner} {et~al.}(1997){Kastner}, {Zuckerman}, {Weintraub}, \&
  {Forveille}}]{kastner1997}
{Kastner}, J.~H., {Zuckerman}, B., {Weintraub}, D.~A., \& {Forveille}, T. 1997,
  Science, 277, 67

\bibitem[{{K\"onigl}(1991)}]{konigl1991}
{K\"onigl}, A. 1991, \apjl, 370, L39

\bibitem[{{Konopacky} {et~al.}(2007){Konopacky}, {Ghez}, {Duch{\^e}ne},
  {McCabe}, \& {Macintosh}}]{konopacky2007}
{Konopacky}, Q.~M., {Ghez}, A.~M., {Duch{\^e}ne}, G., {McCabe}, C., \&
  {Macintosh}, B.~A. 2007, \aj, 133, 2008

\bibitem[{{Lawson} \& {Crause}(2005)}]{lawson2005}
{Lawson}, W.~A., \& {Crause}, L.~A. 2005, \mnras, 357, 1399

\bibitem[{{Lin} {et~al.}(1996){Lin}, {Bodenheimer}, \& {Richardson}}]{lin1996}
{Lin}, D.~N.~C., {Bodenheimer}, P., \& {Richardson}, D.~C. 1996, \nat, 380, 606

\bibitem[{{Lowrance} {et~al.}(2005){Lowrance}, {Becklin}, {Schneider},
  {Kirkpatrick}, {Weinberger}, {Zuckerman}, {Dumas}, {Beuzit}, {Plait},
  {Malumuth}, {Heap}, {Terrile}, \& {Hines}}]{lowrance2005}
{Lowrance}, P.~J., {Becklin}, E.~E., {Schneider}, G., {Kirkpatrick}, J.~D.,
  {Weinberger}, A.~J., {Zuckerman}, B., {Dumas}, C., {Beuzit}, J.-L., {Plait},
  P., {Malumuth}, E., {Heap}, S., {Terrile}, R.~J., \& {Hines}, D.~C. 2005,
  \aj, 130, 1845

\bibitem[{{Luhman} {et~al.}(2003){Luhman}, {Stauffer}, {Muench}, {Rieke},
  {Lada}, {Bouvier}, \& {Lada}}]{luhman2003}
{Luhman}, K.~L., {Stauffer}, J.~R., {Muench}, A.~A., {Rieke}, G.~H., {Lada},
  E.~A., {Bouvier}, J., \& {Lada}, C.~J. 2003, \apj, 593, 1093

\bibitem[{{Macintosh} {et~al.}(2001){Macintosh}, {Max}, {Zuckerman}, {Becklin},
  {Kaisler}, {Lowrance}, {Weinberger}, {Christou}, {Schneider}, \&
  {Acton}}]{macintosh2001}
{Macintosh}, B., {Max}, C., {Zuckerman}, B., {Becklin}, E.~E., {Kaisler}, D.,
  {Lowrance}, P., {Weinberger}, A., {Christou}, J., {Schneider}, G., \&
  {Acton}, S. 2001, in ASP Conf. Ser. 244: Young Stars Near Earth: Progress and
  Prospects, ed. R.~{Jayawardhana} \& T.~{Greene}, 309--+

\bibitem[{{Makarov}(2007)}]{makarov2007}
{Makarov}, V.~V. 2007, \apjs, 169, 105

\bibitem[{{M{\'e}nard} \& {Bertout}(1999)}]{menardbertout1999}
{M{\'e}nard}, F., \& {Bertout}, C. 1999, in NATO ASIC Proc. 540: The Origin of
  Stars and Planetary Systems, ed. C.~J. {Lada} \& N.~D. {Kylafis}, 341--374

\bibitem[{{Montes} \& {Martin}(1998)}]{montes1998}
{Montes}, D., \& {Martin}, E.~L. 1998, \aaps, 128, 485

\bibitem[{{Moss}(2003)}]{moss2003}
{Moss}, D. 2003, \aap, 403, 693

\bibitem[{{Neuh{\"a}user} {et~al.}(2000){Neuh{\"a}user}, {Brandner}, {Eckart},
  {Guenther}, {Alves}, {Ott}, {Hu{\'e}lamo}, \&
  {Fern{\'a}ndez}}]{neuhauser2000}
{Neuh{\"a}user}, R., {Brandner}, W., {Eckart}, A., {Guenther}, E., {Alves}, J.,
  {Ott}, T., {Hu{\'e}lamo}, N., \& {Fern{\'a}ndez}, M. 2000, \aap, 354, L9

\bibitem[{{Paatz} \& {Camenzind}(1996)}]{paatz1996}
{Paatz}, G., \& {Camenzind}, M. 1996, \aap, 308, 77

\bibitem[{{Parker}(1993)}]{parker1993}
{Parker}, E.~N. 1993, \apj, 408, 707

\bibitem[{{Pevtsov} {et~al.}(2003){Pevtsov}, {Fisher}, {Acton}, {Longcope},
  {Johns-Krull}, {Kankelborg}, \& {Metcalf}}]{pevtsov2003}
{Pevtsov}, A.~A., {Fisher}, G.~H., {Acton}, L.~W., {Longcope}, D.~W.,
  {Johns-Krull}, C.~M., {Kankelborg}, C.~C., \& {Metcalf}, T.~R. 2003, \apj,
  598, 1387

\bibitem[{{Piskunov}(1999)}]{piskunov1999}
{Piskunov}, N. 1999, in Astrophysics and Space Science Library, Vol. 243,
  Polarization, ed. K.~N. {Nagendra} \& J.~O. {Stenflo}, 515--525

\bibitem[{{Rajaguru} {et~al.}(2002){Rajaguru}, {Kurucz}, \&
  {Hasan}}]{rajaguru2002}
{Rajaguru}, S.~P., {Kurucz}, R.~L., \& {Hasan}, S.~S. 2002, \apjl, 565, L101

\bibitem[{{Reid}(2003)}]{reid2003}
{Reid}, N. 2003, \mnras, 342, 837

\bibitem[{{Robinson}(1980)}]{robinson1980}
{Robinson}, Jr., R.~D. 1980, \apj, 239, 961

\bibitem[{{Saar}(1988)}]{saar1988}
{Saar}, S.~H. 1988, \apj, 324, 441

\bibitem[{{Saar} \& {Linsky}(1985)}]{saar1985}
{Saar}, S.~H., \& {Linsky}, J.~L. 1985, \apjl, 299, L47

\bibitem[{{Safier}(1999)}]{safier1999}
{Safier}, P.~N. 1999, \apjl, 510, L127

\bibitem[{{Setiawan} {et~al.}(2008){Setiawan}, {Henning}, {Launhardt},
  {M{\"u}ller}, {Weise}, \& {K{\"u}rster}}]{setiawan2008}
{Setiawan}, J., {Henning}, T., {Launhardt}, R., {M{\"u}ller}, A., {Weise}, P.,
  \& {K{\"u}rster}, M. 2008, \nat, 451, 38

\bibitem[{{Shu} {et~al.}(1994){Shu}, {Najita}, {Ostriker}, {Wilkin}, {Ruden},
  \& {Lizano}}]{shu1994}
{Shu}, F., {Najita}, J., {Ostriker}, E., {Wilkin}, F., {Ruden}, S., \&
  {Lizano}, S. 1994, \apj, 429, 781

\bibitem[{{Shu} {et~al.}(1999){Shu}, {Allen}, {Shang}, {Ostriker}, \&
  {Li}}]{shu1999}
{Shu}, F.~H., {Allen}, A., {Shang}, H., {Ostriker}, E.~C., \& {Li}, Z.-Y. 1999,
  in NATO ASIC Proc. 540: The Origin of Stars and Planetary Systems, ed. C.~J.
  {Lada} \& N.~D. {Kylafis}, 193--+

\bibitem[{{Smirnov} {et~al.}(2004){Smirnov}, {Lamzin}, {Fabrika}, \&
  {Chuntonov}}]{smirnov2004}
{Smirnov}, D.~A., {Lamzin}, S.~A., {Fabrika}, S.~N., \& {Chuntonov}, G.~A.
  2004, Astronomy Letters, 30, 456

\bibitem[{{Spruit} \& {Zweibel}(1979)}]{spruit1979}
{Spruit}, H.~C., \& {Zweibel}, E.~G. 1979, \solphys, 62, 15

\bibitem[{{Symington} {et~al.}(2005){Symington}, {Harries}, {Kurosawa}, \&
  {Naylor}}]{symington2005}
{Symington}, N.~H., {Harries}, T.~J., {Kurosawa}, R., \& {Naylor}, T. 2005,
  \mnras, 358, 977

\bibitem[{{Tayler}(1987)}]{tayler1987}
{Tayler}, R.~J. 1987, \mnras, 227, 553

\bibitem[{{Tokunaga} {et~al.}(1990){Tokunaga}, {Toomey}, {Carr}, {Hall}, \&
  {Epps}}]{tokunaga90}
{Tokunaga}, A.~T., {Toomey}, D.~W., {Carr}, J., {Hall}, D.~N.~B., \& {Epps},
  H.~W. 1990, in Instrumentation in astronomy VII; Proceedings of the Meeting,
  Tucson, AZ, Feb. 13-17, 1990 (A91-29601 11-35). Bellingham, WA, Society of
  Photo-Optical Instrumentation Engineers, 1990, p. 131-143., ed. D.~L.
  {Crawford}, 131--143

\bibitem[{{Torres} {et~al.}(2003){Torres}, {Guenther}, {Marschall},
  {Neuh{\"a}user}, {Latham}, \& {Stefanik}}]{torres2003}
{Torres}, G., {Guenther}, E.~W., {Marschall}, L.~A., {Neuh{\"a}user}, R.,
  {Latham}, D.~W., \& {Stefanik}, R.~P. 2003, \aj, 125, 825

\bibitem[{{Valenti} \& {Johns-Krull}(2004)}]{valenti2004}
{Valenti}, J.~A., \& {Johns-Krull}, C.~M. 2004, \apss, 292, 619

\bibitem[{{Valenti} {et~al.}(1995){Valenti}, {Marcy}, \& {Basri}}]{valenti1995}
{Valenti}, J.~A., {Marcy}, G.~W., \& {Basri}, G. 1995, \apj, 439, 939

\bibitem[{{Webb} {et~al.}(1999){Webb}, {Zuckerman}, {Platais}, {Patience},
  {White}, {Schwartz}, \& {McCarthy}}]{webb1999}
{Webb}, R.~A., {Zuckerman}, B., {Platais}, I., {Patience}, J., {White}, R.~J.,
  {Schwartz}, M.~J., \& {McCarthy}, C. 1999, \apjl, 512, L63

\bibitem[{{Weintraub} {et~al.}(2000){Weintraub}, {Saumon}, {Kastner}, \&
  {Forveille}}]{weintraub2000}
{Weintraub}, D.~A., {Saumon}, D., {Kastner}, J.~H., \& {Forveille}, T. 2000,
  \apj, 530, 867

\bibitem[{{White} \& {Hillenbrand}(2004)}]{white2004}
{White}, R.~J., \& {Hillenbrand}, L.~A. 2004, \apj, 616, 998

\bibitem[{{Wichmann} {et~al.}(1998){Wichmann}, {Bastian}, {Krautter},
  {Jankovics}, \& {Rucinski}}]{wichmann1998}
{Wichmann}, R., {Bastian}, U., {Krautter}, J., {Jankovics}, I., \& {Rucinski},
  S.~M. 1998, \mnras, 301, L39+

\bibitem[{{Yang} {et~al.}(2005){Yang}, {Johns-Krull}, \& {Valenti}}]{yang2005}
{Yang}, H., {Johns-Krull}, C.~M., \& {Valenti}, J.~A. 2005, \apj, 635, 466

\bibitem[{{Yang} {et~al.}(2007){Yang}, {Johns-Krull}, \& {Valenti}}]{yang2007}
---. 2007, \aj, 133, 73

\bibitem[{{Zuckerman} {et~al.}(2004){Zuckerman}, {Song}, \&
  {Bessell}}]{zuckerman2004ar}
{Zuckerman}, B., {Song}, I., \& {Bessell}, M.~S. 2004, \apjl, 613, L65

\end{thebibliography}

\clearpage

\begin{table}
  \caption {Journal of Observations}\label{obs}
  \begin{center}
  \leavevmode
  \footnotesize
    \begin{tabular}[h]{lcccc}
    \tableline\tableline
 \ Object  & Wavelength Setting &   UT date    & UT time & Total Exposure Time(s) \\[+5pt]
    \tableline
   TWA 5A  &    2.2228 $\mu$m   & 2000 Jan 08  & 14:08   &    2400         \\[+5pt]
           &    2.2300 $\mu$m   & 2000 Jan 09  & 14:51   &    2400         \\[+5pt]
           &    2.3129 $\mu$m   & 2000 Jan 07  & 15:03   &    2400         \\[+5pt]
   TWA 7   &    2.2228 $\mu$m   & 2000 Jan 08  & 13:22   &    2400         \\[+5pt]
           &    2.2300 $\mu$m   & 2000 Jan 09  & 13:04   &    2400         \\[+5pt]
           &    2.3129 $\mu$m   & 2000 Jan 07  & 14:00   &    3000         \\[+5pt]
   TWA 8A  &    2.2228 $\mu$m   & 2000 Jan 08  & 15:01   &    2400         \\[+5pt]
           &    2.2300 $\mu$m   & 2000 Jan 09  & 13:50   &    2400         \\[+5pt]
           &    2.3129 $\mu$m   & 2000 Jan 11  & 11:43   &    3600         \\[+5pt]
   TWA 9A  &    2.2228 $\mu$m   & 2000 Jan 11  & 11:43   &    3600         \\[+5pt]
           &    2.2300 $\mu$m   & 2000 Jan 12  & 13:26   &    3600         \\[+5pt]
           &    2.3129 $\mu$m   & 2000 Jan 11  & 13:24   &    4800         \\[+5pt]
   TWA 9B  &    2.2228 $\mu$m   & 2000 Jan 11  & 11:43   &    3600         \\[+5pt]
           &    2.2300 $\mu$m   & 2000 Jan 12  & 13:26   &    3600         \\[+5pt]
           &    2.3129 $\mu$m   & 2000 Jan 11  & 13:24   &    4800         \\[+5pt]

    \tableline
  \end{tabular}
 \end{center}
\end{table}

\clearpage
\begin{sidewaystable}
\begin{minipage}{10cm}
  \caption {Magnetic Field Measurements}\label{results}
  \begin{center}
  \leavevmode
  \footnotesize
\resizebox{15.5cm}{!}{
    \begin{tabular}[h]{lccccrrccccccc}
    \tableline\tableline
        &             & $M_*$     & $R_*$     &\Teff& \logg & \vsini & $P_{rot}$\footnote{Taken from \citet{lawson2005}.} & $i$ \footnote{Inclination angles are inferred from rotaion period, stellar radius and \vsini.} &      $L_X$\footnote{Observed X-ray luminosities are derived from \citet{webb1999}.}             & \multicolumn{3}{c}{$\bar B_{obs}$ (kG)}& $B_{eq}$ (kG) \\
 Object &Spectral Type&($M_\odot$)&($R_\odot$)&  (K)&       & (\kms) &  (days) & ($^\circ$)  & ($10^{30}$ ergs $\rm s^{-1})$& M1 Fit ($\chi_r^2$)& M2 Fit ($\chi_r^2$) & M3 Fit ($\chi_r^2$)         &         \\[+5pt]
    \tableline
TWA 5A\footnote{Binary or possible triple system. See discussion in Sec. 3.1.}  & M1.5 & &  & 3700 & 3.70& 38.0 & & & & 4.2 (6.59) & 4.2 (6.49) & 4.9 (6.18) & 1.2  \\[+5pt]
TWA 7   &   M3        & 0.92      & 1.89      & 3300& 3.85  & 9.0    &  5.05  & 28  &       0.875            & 1.6 (1.94)  & 2.0 (1.73)   & 2.3 (1.88)       & 1.3\\[+5pt]   
TWA 8A  &   M2        & 0.60      & 1.29      & 3400& 4.00  & 4.0    &  4.65  & 16  &       0.816            & 2.3 (1.37)  & 2.7 (1.20)   & 3.3 (1.25)       & 1.4\\[+5pt]
TWA 9A  &   K7        & 0.89      & 0.88      & 4000& 4.50  & 9.0    &  5.10  & 90  &       0.679            & 2.4 (2.44)  & 2.9 (2.03)   & 3.5 (1.74)       & 2.1\\[+5pt]
TWA 9B  &   M1        & 0.30      & 0.91      & 3400& 4.00  &10.0    &  3.98  & 60  &                        & 2.8 (1.14)  & 3.3 (1.06)   & 3.1 (1.18)       & 1.4\\[+5pt]
    \tableline
TW Hya\footnote{From \citet{yang2005}.} &   K7        & 2.31      & 0.96      & 4126& 4.84  & 5.8    &  2.80    &  &    1.400            & 2.2   & 2.7   & 2.7       & 2.1\\[+5pt]
    \tableline
    \tableline
  \end{tabular}
}
 \end{center}
 \end{minipage}
\end{sidewaystable}
\thispagestyle{empty}


\clearpage

%

\clearpage

\clearpage

\clearpage

\clearpage

\clearpage

\clearpage

\clearpage

\end{document}